\documentclass[twocolumn,aps,prl,superscriptaddress,floatfix,longbibliography]{revtex4-2}  
\usepackage{chemformula}
\usepackage{graphicx}
\usepackage{dcolumn}
\usepackage{bm}
\usepackage[draft]{changes}
\usepackage[normalem]{ulem}
\usepackage{subfig} 

\usepackage{tabularx}
\usepackage{ragged2e} 
\usepackage{booktabs}

\usepackage{amssymb} 
\usepackage{dcolumn}
\usepackage{bm}
\usepackage[mathlines]{lineno}
\usepackage[colorlinks,linkcolor=blue,anchorcolor=blue,citecolor=blue,urlcolor=blue]{hyperref}
\usepackage{multirow}
\usepackage{color}
\usepackage{amsmath}
\usepackage{afterpage}
\makeatletter
\renewcommand*{\@fnsymbol}[1]{\ensuremath{\ifcase#1\or \dagger\or *\or \ddagger\or
\mathsection\or \mathparagraph\or \|\or **\or \dagger\dagger \or
\ddagger\ddagger \else\@ctrerr\fi}} \makeatother

\usepackage{cleveref}  
  
\usepackage{caption}  
\begin{document}
\title{The Impact of Ionic Anharmonicity on Superconductivity in Metal-Stuffed B-C Clathrates}

\author{Wenbo Zhao}
\thanks{These two authors contributed equally}
\affiliation{Key Laboratory of Material Simulation Methods and Software of Ministry of Education, College of Physics, Jilin University, Changchun 130012, China}
\affiliation{International Center of Future Science, Jilin University, Changchun 130012, China} 

\author{Ying Sun}
\thanks{These two authors contributed equally}
\affiliation{Key Laboratory of Material Simulation Methods and Software of Ministry of Education, College of Physics, Jilin University, Changchun 130012, China}
\affiliation{International Center of Future Science, Jilin University, Changchun 130012, China}

\author{Jiaxiang Li}
\affiliation{Key Laboratory of Material Simulation Methods and Software of Ministry of Education, College of Physics, Jilin University, Changchun 130012, China}

\author{Peng Yuan}
\affiliation{College of Physics, Jilin University, Changchun 130012, China}

\author{Toshiaki Iitaka}
\affiliation{Discrete Event Simulation Research Team, RIKEN Center for Computational Science, 2-1 Hirosawa, Wako, Saitama, 351-0198, Japan}

\author{Xin Zhong}
\affiliation{Key Laboratory of Material Simulation Methods and Software of Ministry of Education, College of Physics, Jilin University, Changchun 130012, China}

\author{Hefei Li}
\email{lihefei37@jlu.edu.cn}
\affiliation{Key Laboratory of Material Simulation Methods and Software of Ministry of Education, College of Physics, Jilin University, Changchun 130012, China}
\affiliation{State Key Laboratory of Superhard Materials, College of Physics, Jilin University, Changchun 130012, China}

\author{Yue-Wen Fang}
\email{yuewen.fang@ehu.eus}
\affiliation{Fisika Aplikatua Saila, Gipuzkoako Ingeniaritza Eskola, University of the Basque Country (UPV/EHU), Europa Plaza 1, 20018 Donostia/San Sebasti{\'a}n, Spain}
\affiliation{Centro de F{\'i}sica de Materiales (CFM-MPC), CSIC-UPV/EHU, Manuel de Lardizabal Pasealekua 5, 20018 Donostia/San Sebasti{\'a}n, Spain}	

\author{Hanyu Liu}
\email{hanyuliu@jlu.edu.cn}
\affiliation{Key Laboratory of Material Simulation Methods and Software of Ministry of Education, College of Physics, Jilin University, Changchun 130012, China}
\affiliation{International Center of Future Science, Jilin University, Changchun 130012, China}

\author{Ion Errea}
\affiliation{Fisika Aplikatua Saila, Gipuzkoako Ingeniaritza Eskola, University of the Basque Country (UPV/EHU), Europa Plaza 1, 20018 Donostia/San Sebasti{\'a}n, Spain}
\affiliation{Centro de F{\'i}sica de Materiales (CFM-MPC), CSIC-UPV/EHU, Manuel de Lardizabal Pasealekua 5, 20018 Donostia/San Sebasti{\'a}n, Spain}
\affiliation{Donostia International Physics Center (DIPC), Manuel de Lardizabal Pasealekua 4, 20018 Donostia/San Sebasti{\'a}n, Spain}

\author{Yu Xie}
\email{xieyu@jlu.edu.cn}
\affiliation{Key Laboratory of Material Simulation Methods and Software of Ministry of Education, College of Physics, Jilin University, Changchun 130012, China}
\affiliation{Key Laboratory of Physics and Technology for Advanced Batteries of Ministry of Education, College of Physics, Jilin University, Changchun 130012, China}

\date{\today}
\begin{abstract}
\noindent\textbf{Abstract:}~Metal-stuffed B$-$C compounds with sodalite clathrate structure have captured increasing attention due to their predicted exceptional superconductivity above liquid nitrogen temperature at ambient pressure. However, by neglecting the quantum lattice anharmonicity, the existing studies may result in an incomplete understanding of such a lightweight system. Here, using state-of-the-art \textit{ab initio} methods incorporating quantum effects and machine learning potentials, we revisit the properties of a series of $XY$$\text{B}_{6}\text{C}_{6}$ clathrates where $X$ and $Y$ are metals. Our findings show that ionic quantum and anharmonic effects can harden the $E_g$ and $E_u$ vibrational modes, enabling the dynamical stability of 15 materials previously considered unstable in the harmonic approximation, including materials with previously unreported ($XY$)$^{1+}$ state, which is demonstrated here to be crucial to reach high critical temperatures. Further calculations based on the anisotropic Migdal-Eliashberg equation demonstrate that the $T_\text{c}$ values for KRb$\text{B}_{6}\text{C}_{6}$ and Rb$\text{B}_{3}\text{C}_{3}$ among these stabilized compounds are 102 and 115 K at 0 and 15 GPa, respectively, both being higher than $T_\text{c}$ of 92 K of KPb$\text{B}_{6}\text{C}_{6}$ at the anharmonic level. These record-high $T_\text{c}$ values, surpassing liquid nitrogen temperatures, emphasize the importance of anharmonic effects in stabilizing B–C clathrates with large electron-phonon coupling strength and advancing the search for high-$T_\text{c}$ superconductivity at (near) ambient pressure.
\end{abstract}
\maketitle

\section{I. Introduction}
The quest for high-temperature superconductivity has predominantly centered on unconventional superconductors for nearly four decades following the discovery of cuprates. However, recent theoretical-driven experimental observations of superconducting critical temperature $T_\text{c}$ $>$ 200 K  in superhydrides gradually refocus the attention to conventional superconductors\cite{Overview-2021,OverviewofSuperconductivityYing-1,OverviewofSuperconductivityYing-2,Overview-Cui}. In these materials, electrons are coupled via phonons, as elucidated by the Bardeen-Cooper-Schrieffer (BCS) theory\cite{BCS-1,BCS-2}, which offers an excellent platform for the theoretical design of high-temperature superconductors. Unfortunately, experimental realization of high-temperature superconductivity in superhydrides typically requires high pressures exceeding 150 GPa, which strictly forbids their practical applications\cite{OverviewofSuperconductivityYing-1}. Despite significant efforts to design high-$T_\text{c}$ superhydrides at reduced pressures, none of these proposed compounds have been realized experimentally below 50 GPa\cite{OverviewofSuperconductivityYing-1}. Thus, searching for synthesizable superconductors with high-$T_\text{c}$ at low or close to ambient pressure remains an immediate yet challenging task.

Light-element-based covalent metals were regarded as potential high-$T_\text{c}$ superconductors at near ambient pressure because the presence of metallic covalent bonds could lead to large phonon frequencies, significant electron-phonon coupling (EPC)\cite{Belli-2021}, and high density of the states at Fermi level ($N$($E_{f}$)), which are the most critical factors for conventional superconductivity\cite{BCS-1,BCS-2}. MgB$_{2}$ with a $T_\text{c}$ of 39 K\cite{MgB2-original}, resulting from a strong coupling of $\sigma$-bonding electrons with in-plane B$-$B vibrational phonons, is the best example of this concept, which holds the record of $T_\text{c}$ among the conventional superconductors at ambient pressure. Other covalent superconductors with lower $T_\text{c}$s are also observed experimentally\cite{review-BandCClathrates,SiandC-clathrates,B-QC,metal-intercalated-graphites2015}. Based on this principle, several high-$T_\text{c}$ superconductor candidates were proposed, including heavily boron-doped diamond\cite{heavily-boron-doped-diamond}, boron-doped graphane\cite{boron-doped-graphane}, doped carbon clathrates\cite{doped-C-clathrates-77k}, layered metal borocarbides\cite{layered-metal-intercalated-borocarbides-Li,layered-metal-intercalated-borocarbides-M}, B$-$C compounds\cite{BC5,boron-carbon-Material-trends}, and Sr$\text{B}_{3}\text{C}_{3}$\cite{srbc-prr,srbc-sci}. Remarkably, boron appears in almost all materials, serving as a hole dopant to metalize the filled $\sigma$ carbon bonds while maintaining large EPC and high phonon modes through boron-carbon bonding.

A remarkable recent achievement in the field is that Sr$\text{B}_{3}\text{C}_{3}$ has been successfully synthesized at 57 GPa and 2500 K\cite{srbc-sci}, and the predicted superconductivity is also confirmed with an onset $T_\text{c}$ of 20 K at 40 GPa\cite{srbc-sci,srbc-prr}. Superconductivity originates from the interplay between $\sigma$ B$-$C bonds and boron $E_g$ phonon modes\cite{SrBC-BaBC,srbc-prr,srbc-sci}. The thermodynamically stable range of Sr$\text{B}_{3}\text{C}_{3}$ is from 50 to 200 GPa\cite{srbc-sci,srbc-prr}, but there is experimental evidence that it may survive close to ambient pressure, where its superconducting $T_\text{c}$ approximates 41 K\cite{srbc-sci}, comparable to MgB$_{2}$\cite{MgB2-original}. Notably, Sr$\text{B}_{3}\text{C}_{3}$ resembles a host-guest structure, where the metal atom is encapsulated within the B$-$C cage, offering an excellent opportunity to adjust the properties by varying the stuffed metal atom\cite{srbc-sci,srbc-prr,Rb0.4,SrRb,XBC,XYBC,BC-NH4}. The subsequently synthesized insulating La$\text{B}_{3}\text{C}_{3}$\cite{LaBC} has confirmed that for metal-stuffed B$-$C clathrates, the shape of the density of states (DOS) remains unchanged, and the metal atom's valence state determines the Fermi level. Therefore, adjusting the metal's valence state to position the Fermi level at the DOS peaks can further enhance the superconductivity of metal-stuffed B$-$C clathrates.

It has been predicted that the superconducting critical temperature can be further enhanced by substituting Sr$^{2+}$ with two metal atoms $X$ and $Y$ in 1$:$1  ratio, with many different valence states (($XY$)$^{n+}$ = \(\frac{X^{m+}+Y^{l+}}{2}\))\cite{SrRb,XBC,XYBC,Rb0.4}. Most of the designed high-$T_\text{c}$ $XY$$\text{B}_{6}\text{C}_{6}$ compounds have an average metal valence state of +1.5, where KPb$\text{B}_{6}\text{C}_{6}$ currently shows the highest predicted $T_\text{c}$ of 88 K\cite{XYBC}. Although the $T_\text{c}$ has been doubled in comparison with that of MgB$_{2}$, many $XY$$\text{B}_{6}\text{C}_{6}$ compounds with a Fermi level positioned at the DOS peak, particularly those with an average metal valence state of +1, were predicted to be dynamically unstable. However, the light-weight nature of $XY$$\text{B}_{6}\text{C}_{6}$ compounds suggests that quantum lattice anharmonicity should play a crucial role in determining their dynamic and superconducting properties\cite{SSCHA-H3S-1,SSCHA-LaH10,BC-NH4}, because the standard harmonic approximation may not adequately represent the Born-Oppenheimer energy surface in the range defined by the zero point energy. Thus, to gain a unified understanding of the superconductivity in metal-stuffed B$-$C clathrates, it is an urgent and necessary task to investigate these compounds under the inclusion of anharmonic effects.

Here, we re-examine the dynamical stability and superconductivity of metal-stuffed B$-$C  clathrates ($X$$^{n+}$[$\text{B}_{3}\text{C}_{3}$]$^{n-}$, $X$ can be a single metal or two metals, $n$ denotes the average valence state) by including quantum lattice anharmonic effects within the stochastic self-consistent harmonic approximation (SSCHA)\cite{SSCHA-program,sscha-2013,SSCHA-2014,sscha-2017,sscha-2018}. Meanwhile, an attention-centered neural network was employed to construct machine learning potentials to address the high computational cost of SSCHA. Our results reveal that anharmonic effects can modify the stability of metal-stuffed B$-$C  clathrates with different metal valence states, allowing us to identify 15 new stable clathrate compounds, including 3 $X$$^{1+}$[$\text{B}_{3}\text{C}_{3}$]$^{1-}$ compounds. Our in-depth analysis shows that the average valence state of plus one is the most convenient for high-temperature superconductivity. By utilizing the anisotropic Migdal-Eliashberg formalism\cite{eliashberg-1960,eliashberg-2007,eliashberg-2018}, we confirm $X$$^{1+}$[$\text{B}_{3}\text{C}_{3}$]$^{1-}$ compounds to have a higher $T_\text{c}$ than that of $X$$^{1.5+}$[$\text{B}_{3}\text{C}_{3}$]$^{1.5-}$ materials. The predicted $T_\text{c}$ is 102 K for KRb$\text{B}_{6}\text{C}_{6}$ at 0 GPa and 115 K for Rb$\text{B}_{3}\text{C}_{3}$ at 15 GPa. This is a significant increase of $\sim$ 25 \% compared to the 92 K estimated in KPb$\text{B}_{6}\text{C}_{6}$ at the anharmonic level, establishing the highest possible $T_\text{c}$ value for metal-stuffed B$-$C clathrates.

\section{II. Results}
\subsection{Electronic Properties}
\begin{figure*}[t]
    \includegraphics[width=2\columnwidth]{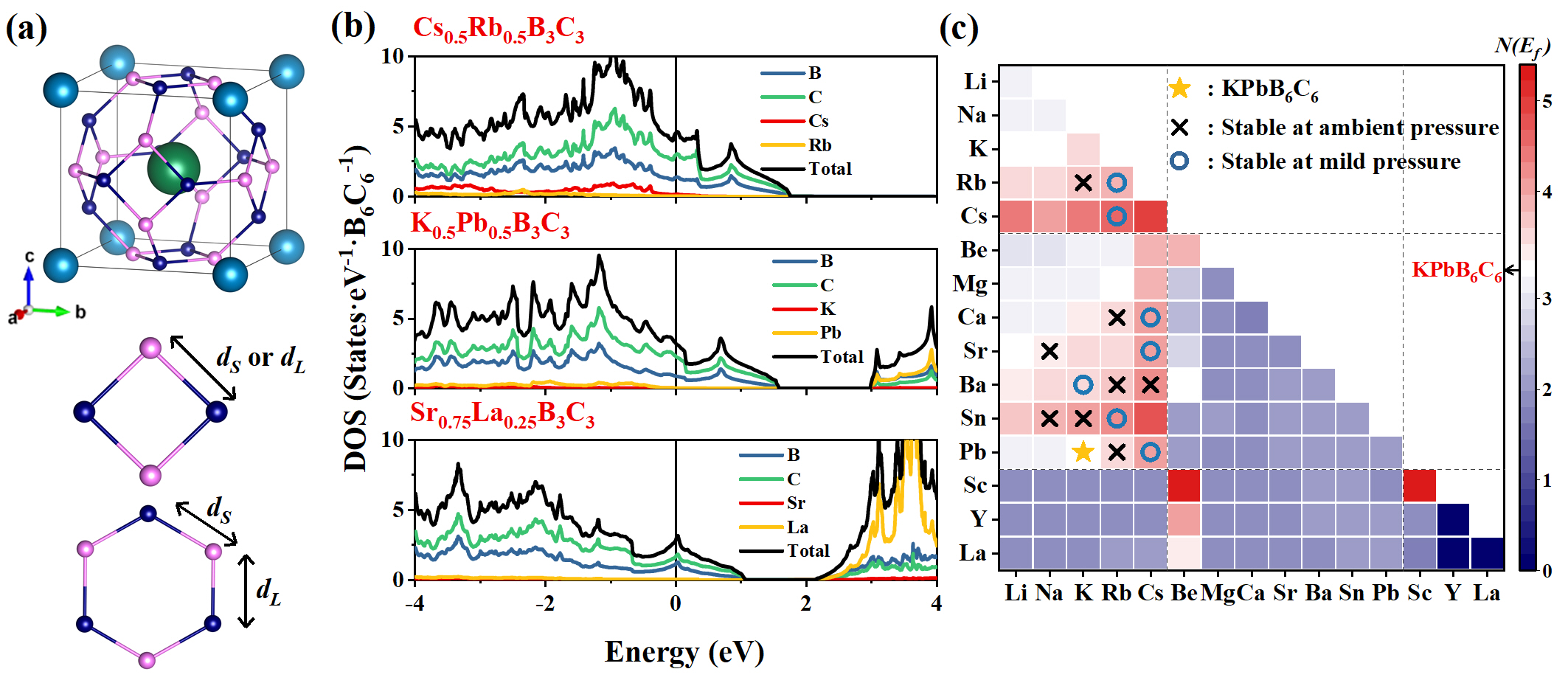}
    \caption{(a) Crystal structure of $XY$$\text{B}_{6}\text{C}_{6}$, where light blue and green spheres represent metal atoms ($X$ and $Y$), and pink and dark blue spheres represent B and C atoms, respectively. (b) Projected electronic density of states (PDOS) for the \ch{Cs0_.5Rb0_.5B3C3}, \ch{K0_.5Pb0_.5B3C3} and \ch{Sr0_.75La0_.25B3C3}\cite{Rb0.4}. Vertical lines indicate the Fermi level. (c) The heat map in lower triangle shows the density of states at the Fermi level, $N$($E_{f}$), in units of states$\cdot$eV$^{-1}$$\cdot$$\text{B}_{6}\text{C}_{6}$$^{-1}$ for various $XY$$\text{B}_{6}\text{C}_{6}$ combinations at ambient pressure. Metals $X$ and $Y$ are on the x- and y-axes. The color scale is centered on KPb$\text{B}{_6}\text{C}{_6}$ (yellow star), shown in white. Combinations with higher $N$($E_{f}$) are red, and those with lower values are blue. Black crosses indicate dynamically stable combinations at ambient pressure considering anharmonic effects, and pink hexagons show stability under mild pressure (20 GPa). Black dashed lines separate combinations based on the average valence state of the metals.
    \label{Fig1}}
\end{figure*}
We first examine the DOS of $XY$$\text{B}_{6}\text{C}_{6}$ compounds as shown in Figure \ref{Fig1}. The DOS at the Fermi level ($N$($E_{f}$)) is primarily contributed by the B$-$C framework, with negligible impact from the metal atoms as observed in previous reports\cite{XYBC,XBC}. Different metal atom combinations position the Fermi level differently. When the average valence state of metal atoms is +1, +1.5, or +2.25, the Fermi level aligns with DOS peaks (Figure \ref{Fig1}b), indicating the potential for high-$T_\text{c}$ superconductivity. A detailed analysis of charge transfer from the metal atoms to the B--C framework, based on Löwdin schemes, is provided in Supplementary Note~7\cite{Bader,COHP-0,COHP-1,Mulliken,lobster}. We need to mention that +2.25 average metal state can only be achieved in $X_{3}Y$B$_{12}$C$_{12}$ compounds\cite{Rb0.4} with a larger supercell. Moreover, the height of the +2.25 DOS peak is almost the same as the +1.5 DOS peak and lower than the +1 DOS peak. Thus, we focus on the properties of $XY$$\text{B}_{6}\text{C}_{6}$ compounds with +1 and +1.5 average metal states in the following discussion. The $N$($E_{f}$) heat map of $XY$$\text{B}_{6}\text{C}_{6}$ against metal combinations (Figure \ref{Fig1}c) shows clear partitioning due to different metal average valence state. Generally, the $N$($E_{f}$) follows the trend +1 $>$ +1.5 $>$ +2 $\approx$ +2.5 with few exceptions, where metal atom Be or Sc is the main contributor to the $N$($E_{f}$). Comparing with the previously reported highest $N$($E_{f}$) of KPb$\text{B}_{6}\text{C}_{6}$ (marked as the yellow star), we identify 34 $XY$$\text{B}_{6}\text{C}_{6}$ compounds with larger $N$($E_{f}$), where all of them were predicted to be dynamically unstable at ambient pressure within the harmonic approximation.  

\begin{figure}[t]
    \includegraphics[width=1\columnwidth]{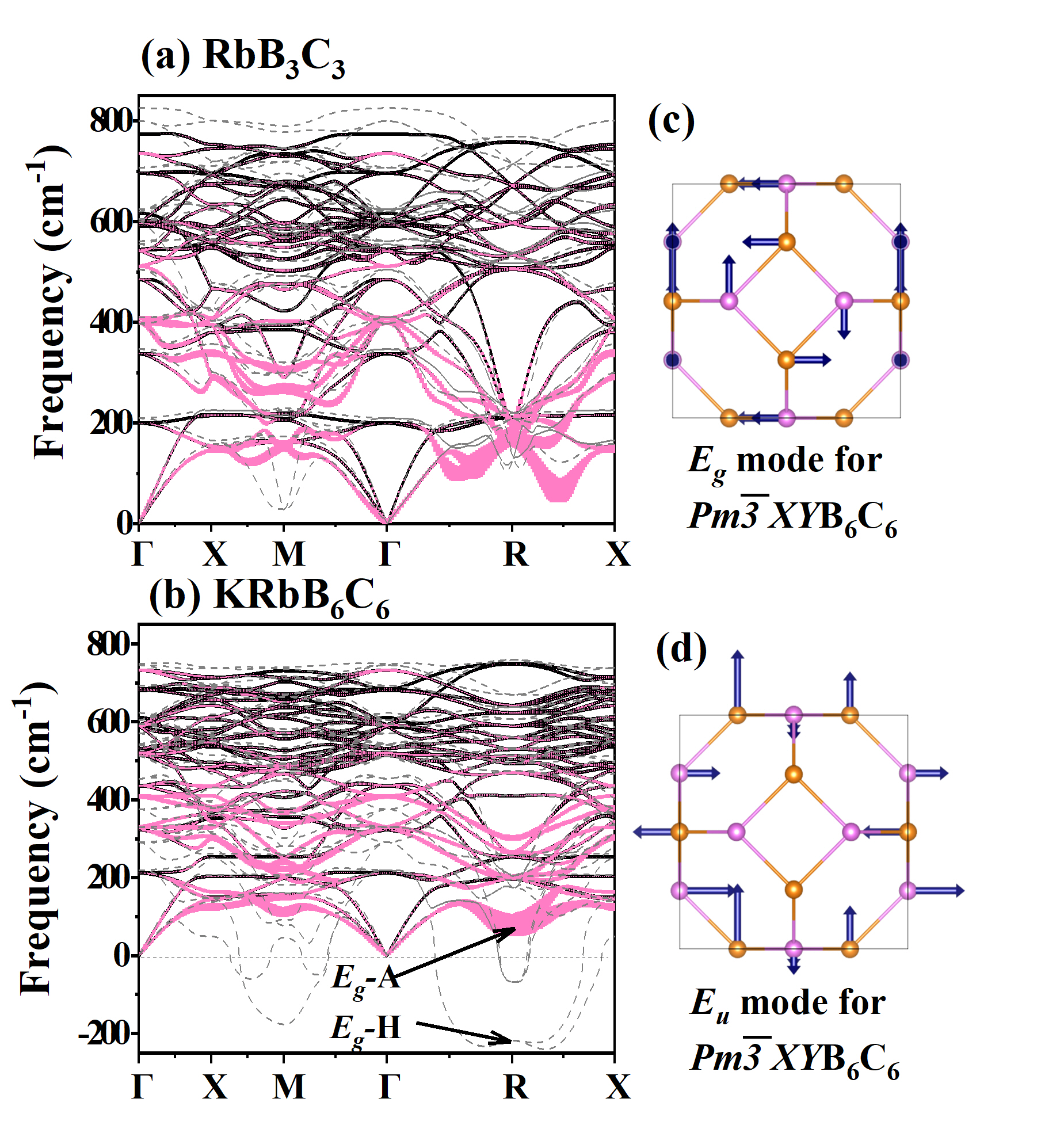}
    \flushleft
    \caption{The gray dashed line represents the harmonic spectrum, and the black solid line represents the anharmonic spectrum obtained from the 4-order Hessian matrix, where pink solid circles indicate the electron-phonon coupling strength, with the radius proportional to its strength for (a) RbB$_3$C$_3$ and (b) KRbB$_6$C$_6$. (c) and (d) show the visualized vibrational eigenmodes of the $E_g$ and $E_u$ phonon modes, respectively.
    \label{Fig2}}
\end{figure}

\subsection{Anharmonic Lattice Dynamics and Stability}
Given the low atomic masses of boron and carbon, quantum lattice anharmonic effects might be significant enough to modulate the dynamical properties of $XY$$\text{B}_{6}\text{C}_{6}$. We then employ the stochastic self-consistent harmonic approximation (SSCHA) method\cite{sscha-2013,SSCHA-2014,sscha-2017,sscha-2018} to evaluate the lattice parameter, dynamic stability and superconducting properties of the 34 $XY$$\text{B}_{6}\text{C}_{6}$ compounds with the higher $N$($E_{f}$), where Sr$\text{B}_{3}\text{C}_{3}$ and KPb$\text{B}_{6}\text{C}_{6}$ are also considered for better comparison. 

Notably, the SSCHA method is a rigorous variational approach that directly obtains the anharmonic dynamical matrix by minimizing the quantum free energy functional, achieved stochastically via Monte Carlo summation and importance sampling\cite{Annealed-importance-sampling2001,Monte-Carlo-1994} over several consecutive ensembles (populations) of a large number of individuals ($N_\text{c}$)\cite{sscha-2013,SSCHA-2014,sscha-2017,sscha-2018}. However, the method is computationally demanding, particularly for slowly converging soft modes, which often require multiple iterations with population sizes ranging from tens to hundreds of thousands. To address this issue, we develop an SSCHA-ACNN (attention-centered neural network) workflow that employs machine learning potentials (MLPs) and active learning to accelerate the simulations. The accuracy of the SSCHA-ACNN workflow is validated by root-mean-square errors below 0.5 meV$\cdot$atom$^{-1}$ for total energies and 50 meV$\cdot$\AA$^{-1}$ for forces (Supplementary Figures~2$-$3), along with excellent agreement in phonon spectra compared to direct DFT-based SSCHA calculations (Supplementary Figure~4).

The SSCHA-ACNN workflow begins with DFT-level calculations to refine the anharmonic energy surface (AES) and generate a dataset for subsequent MLP-level calculations. At this stage, 8 materials exhibit phonon dispersions without imaginary frequencies (Supplementary Figures~22$-$23), suggesting that they become dynamically stable at ambient pressure due to the inclusion of quantum lattice anharmonic effects. For the remaining materials, we proceed with MLP-based calculations, increasing the number of individual configurations in the ensemble to over 10,000, and find that they are all dynamically unstable at ambient pressure. Among these, 7 materials exhibit stable auxiliary SSCHA dynamical matrices, but their Hessian dynamical matrices still show imaginary frequencies, indicating that the ${Pm\bar{3}}$ structure is not a local minimum in the AES. We further analyzed their soft phonon modes to explore possible symmetry-breaking distortions, but all resulting distorted structures remain unstable (Supplementary Note~8). For the other 18 materials, phonon frequency changes during the SSCHA simulations show that some optical phonon frequencies rapidly drop to zero and remain constant as the ensemble size grows (Supplementary Figures~7$-$16). This behavior suggests flat wave functions and weak restoring forces, reflecting the instability of the ${Pm\bar{3}}$ structure and a shallow AES with many energetically comparable local minima and nearly barrierless transition paths. Since high pressure is an effective way to tune the dynamic stability of a compound, we conduct additional SSCHA simulations up to 20 GPa, which can be easily achieved in the large-volume press, resulting in 7 more stable compounds (Supplementary Figure~24).

Since AES is changed upon introducing quantum lattice anharmonic effects, the crystal structure of materials can be modified. The prototype structure of $XY$$\text{B}_{6}\text{C}_{6}$ is based on 2Sr@$\text{B}_{6}\text{C}_{6}$ (Sr$\text{B}_{3}\text{C}_{3}$), which has a type-VII clathrate framework\cite{srbc-sci}. This host framework consists of truncated octahedral cages with six four-sided and eight six-sided faces ($4^66^8$). Each cage comprises 24 vertices, alternating between C and B atoms, with an $X$ or $Y$ guest cation at the center. When $X$ and $Y$ are identical cations ($X$$\text{B}_{6}\text{C}_{6}$), all neighboring B and C atoms have equal bond lengths ($d_\text{B$-$C}$). The anharmonic effect solely increases the lattice parameters of the structure. However, when $X$ and $Y$ are different cations, their differences in electric properties and ionic radii create distinct chemical environments around the boron and carbon atoms, resulting in two unequal B$-$C bond lengths: a longer bond, $d_\text{L}$, and a shorter bond, $d_\text{S}$. Our results indicate that anharmonic effects do not alter the lattice symmetry but do increase the lattice parameters and slightly shift the positions of B and C atoms, and enlarge the difference between $d_\text{L}$ and $d_\text{S}$.

The changes in crystal structure lead to a substantial renormalization of the phonon spectrum at the anharmonic level (Figure~\ref{Fig2} a,b), particularly evident as an overall softening of the phonon with increasing lattice constant, as well as the softening or hardening of phonon vibrational modes --- especially the $E_g$ and $E_u$ modes (Figure~\ref{Fig2} c,d) ---corresponding to the expansion or contraction of B$-$C bond lengths. The $E_g$ mode involves the libration of two boron and two carbon atoms located at opposite corners of the square faces, resulting in the distortion of the quadrilateral into a rectangle, with two B$-$C edges elongated and two shortened. The $E_u$ mode features oscillation of the same pairs of boron and carbon atoms at opposite corners, where two atoms move upward and two move downward, altering the symmetry of the face. Both the $E_g$ and $E_u$ modes tend to distort the square faces of the B$-$C cages and are key contributors to the instability in $XY$$\text{B}{_6}\text{C}{_6}$ at the harmonic level (Supplementary Figures~22$-$24). 

Anharmonic effects results in longer $d_\text{B$-$C}$ of $X$$\text{B}_{3}\text{C}_{3}$ and the expansion in lattice constant, which softens nearly all phonon modes, especially the acoustic branches near the R point, contributing significantly to the electron-phonon coupling (EPC) strength. In contrast, for $XY$$\text{B}_{6}\text{C}_{6}$, anharmonic effects increase the difference between $d_\text{L}$ and $d_\text{S}$. The decrease in $d_\text{S}$ significantly hardens the $E_g$ and $E_u$ modes, making eight $XY$$\text{B}_{6}\text{C}_{6}$ materials dynamically stable at ambient pressure. However, another eight compounds remain unstable because the impact of anharmonic corrections on enhancing the overall stability of certain materials is limited. Among the eight structures that are unstable under ambient pressure and anharmonic effects, seven of them can be stabilized at moderate pressures, as high pressure can reduce the lattice constant and compress the B$-$C bond length. This compression hardens the phonon modes, thereby enhancing the stability of crystal structures.

\subsection{Superconducting Properties}
\begin{figure}[t]
    \includegraphics[width=1\columnwidth]{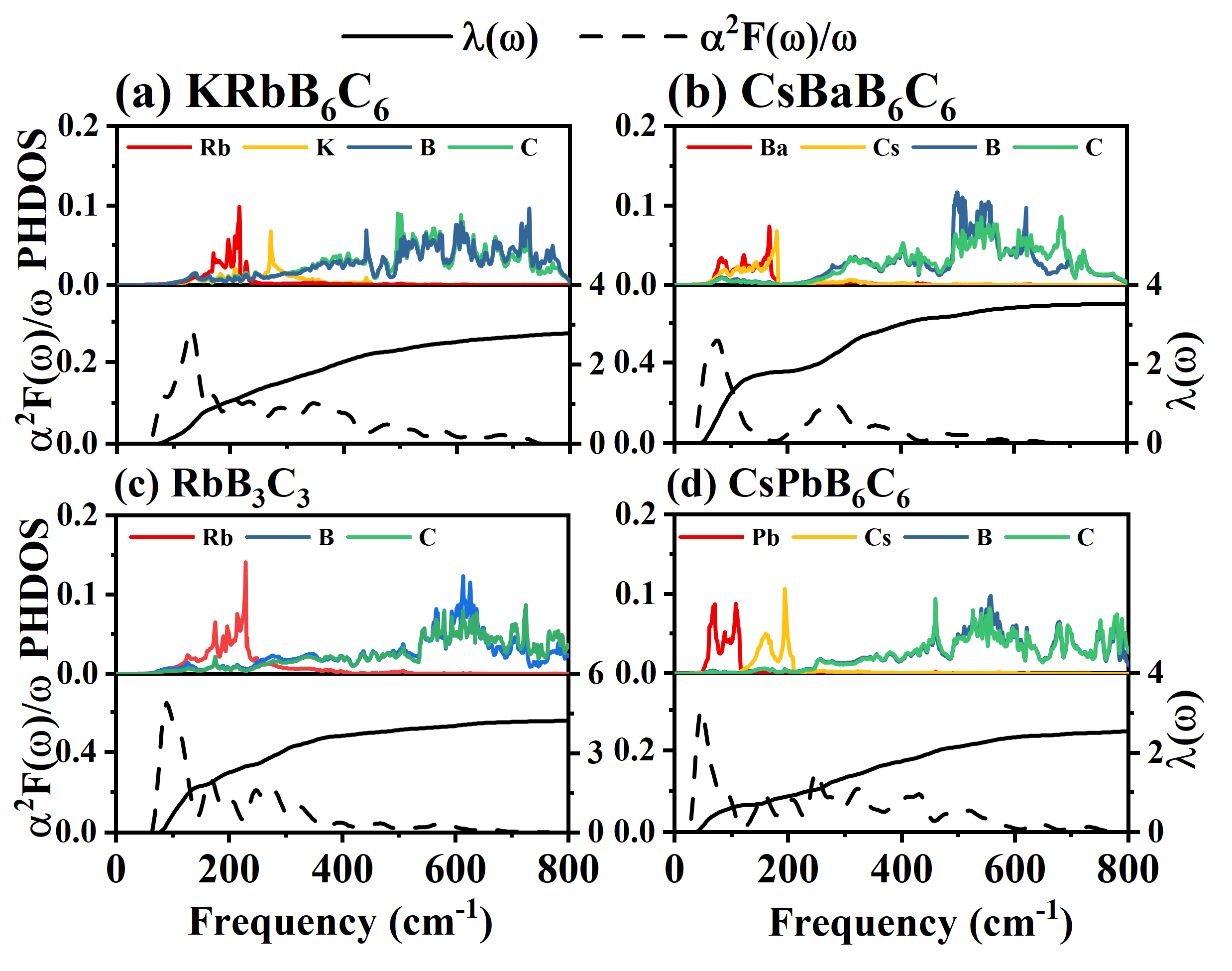}
    \flushleft
    \caption{Projected phonon density of states (PHDOS), Eliashberg spectral function $\alpha^2F(\omega)/\omega$ and integral $\lambda(\omega)$ for (a) KRb$\text{B}_{6}\text{C}_{6}$, (b) CsBa$\text{B}_{6}\text{C}_{6}$, (c) Rb$\text{B}_{3}\text{C}_{3}$, (d) CsPb$\text{B}_{6}\text{C}_{6}$.
    \label{Fig3}}
\end{figure}

According to Eliashberg theory, phonon softening may enhance the electron-phonon coupling, favoring a high $T_\text{c}$. We calculated the superconducting critical temperatures using the isotropic Eliashberg equations. Anharmonic effects soften the phonon vibration modes of $X$$\text{B}_{3}\text{C}_{3}$, so their anharmonic $T_\text{c}$ is usually higher than the harmonic one. For example, the estimated $T_\text{c}$ of Sr$\text{B}_{3}\text{C}_{3}$ is renormalized from 27 K (harmonic approximation) to 34 K (anharmonic effects), and that of Rb$\text{B}_{3}\text{C}_{3}$ from 94 K to 98 K. For $XY$$\text{B}_{6}\text{C}_{6}$, it's more complex. Anharmonic effects soften the acoustic-branch vibration modes related to the lattice constant and harden the $E_{g}$ and $E_{u}$ vibration modes of the B$-$C bonds with length $d_{S}$. The latter contributes more to the electron-phonon coupling interaction, making the anharmonic $T_\text{c}$ generally lower than the harmonic one. For instance, in KPbB$_6$C$_6$, the $E_g$ and $E_u$ modes harden from 113 to 240cm $^{-1}$ and from 180 to 269 cm$^{-1}$, respectively, and their contribution to $\lambda$ correspondingly decreases from 36\% to 20\%. The theoretical $T_\text{c}$ of KPbB$_6$C$_6$ is thus renormalized from 88 K to 77 K. Supplementary Note~4 compares phonon spectra from third- and fourth-order calculations, confirms the negligible effect of fourth-order terms, and provides a more detailed analysis of harmonic and anharmonic phonon modes and their impact on superconductivity. As previous studies have suggested significant anisotropy\cite{SrBC-BaBC}, we also solved the fully anisotropic Eliashberg equations for representative high-$T_\text{c}$ compounds. The anharmonic $T_\text{c}$ reaches 102 K in KRbB$_6$C$_6$ at ambient pressure and 115 K in RbB$_3$C$_3$ at 15 GPa—25\% higher than the 92 K in KPbB$_6$C$_6$ at the same level. 

We also analyzed the superconducting mechanisms of 15 newly stabilized structures (Figure~3 and Supplementary Figures~22$-$24) --- 8 stable at ambient pressure and 7 at moderate pressure --- under anharmonic effects. In most studied $XY$B$_6$C$_6$ compounds, strong electron-phonon coupling arises from softened acoustic modes and selected optical modes, particularly $E_g$ and $E_u$, which are critical for enhancing superconductivity. By contrast, compounds such as SrNaB$_6$C$_6$ and CaRbB$_6$C$_6$, which lack softened acoustic modes, exhibit significantly lower $T_\text{c}$ values (see Supplementary Table~5).

We compared the average valence state of metal atoms, density of states at the Fermi level $N$($E_{f}$), EPC constant $\lambda$, logarithmic average phonon frequency $\omega_{\text{log}}$, and $T_\text{c}$ values (Figure \ref{Fig4} and see Supplementary Table~5). Our results show that there is no clear linear relationship between $N$($E_{f}$) and $T_\text{c}$. As analyzed in our previous work, the final value of $T_\text{c}$ results from a complex interplay among a high density of states, appropriate phonon energies, and Coulomb repulsion\cite{Tiago-AFM2024}. Instead, the average valence state of metal atoms significantly influences $T_\text{c}$: compounds with an average valence state of +1 have higher $T_\text{c}$ values than those with +1.5, and materials with the same valence state exhibit similar $T_\text{c}$ ranges due to adjustments in the Fermi level. Valence state of +1 and +1.5 shift the Fermi level to different peaks on either side of a shoulder in the DOS, with the +1 peak slightly higher than the +1.5 peak, explaining the higher $T_\text{c}$ for +1 compounds. Variations among metal atoms lead to slight differences in peak values for $XY$$\text{B}_{6}\text{C}_{6}$ compounds, making these peaks numerically close and hard to distinguish. 

\begin{figure}[t]
    \includegraphics[width=1\columnwidth]{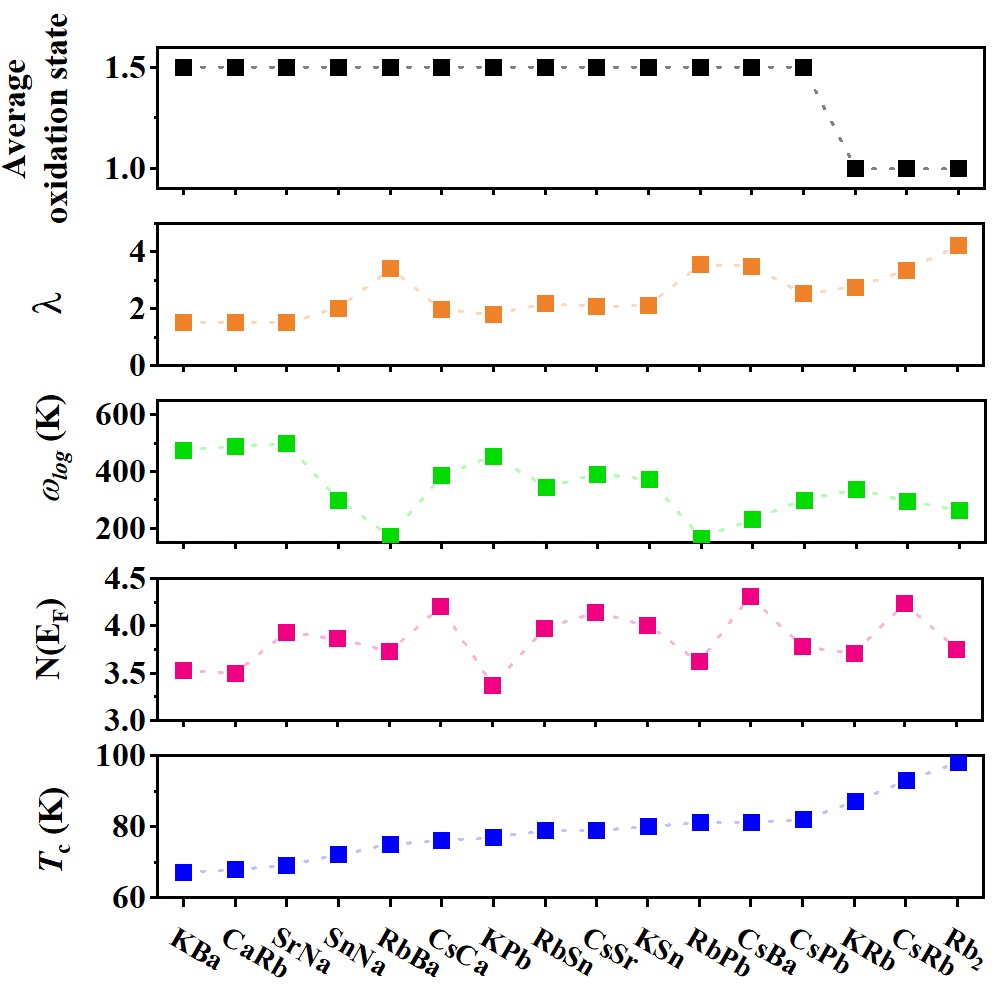}
    \flushleft
    \caption{The average valence state of metal atoms, density of states at the Fermi level $N$($E_{f}$), EPC constant $\lambda$, phonon frequency logarithmic average $\omega_{\text{log}}$, and superconducting critical temperature $T_\text{c}$ for $XY$$\text{B}_{6}\text{C}_{6}$ compounds.
    \label{Fig4}}
\end{figure}

\subsection{Experimental Synthesizability}
Thermodynamic stability analysis was performed to estimate the synthesis conditions for high-temperature superconducting $XY$B$_6$C$_6$ compounds, similar to those used for SrB$_3$C$_3$ and LaB$_3$C$_3$, which can be quenched to ambient conditions after synthesis at mild pressures\cite{LaBC,srbc-sci,srbc-prr}. Free energy calculations, incorporating decomposition enthalpy relative to the ground-state elemental structures and configurational entropy, were used to provide an upper limit for the pressure and temperature conditions required for $XY$B$_6$C$_6$ compounds to overcome the enthalpic penalty, same as Geng et al.\cite{XYBC}. The results indicate that all predicted $XY$B$_6$C$_6$ compounds—except RbSnB$_6$C$_6$, CsRbB$_6$C$_6$, CsPbB$_6$C$_6$, and RbPbB$_6$C$_6$—are thermodynamically stable under conditions accessible to laser-heated diamond anvil cell (DAC) experiments (see Supplementary Table~4). For example, SrNaB$_6$C$_6$, CaRbB$_6$C$_6$, and KBaB$_6$C$_6$ are stabilized at 0 K and pressures below 30 GPa. Notably, KRbB$_6$C$_6$ (50 GPa/1000 K) and RbB$_3$C$_3$ (30 GPa/4000 K), which exhibit the highest predicted $T_\text{c}$ values, are stabilizable under conditions comparable to those used in the synthesis of SrB$_3$C$_3$ (55–60 GPa, 3000 K)\cite{srbc-sci,DAC} (see Supplementary Note~5).

\section{III. DISCUSSION}
Using state-of-the-art \textit{ab initio} methods incorporating quantum effects and machine learning potentials, we systematically investigated the dynamical stability and superconducting properties of B$-$C clathrate compounds. Fifteen new dynamically stable materials with large electron-phonon coupling constants are identified, including previously unreported materials with +1 average valence state, where quantum fluctuations are essential to sustain their structures. Notably, KRb$\text{B}_{6}\text{C}_{6}$ and Rb$\text{B}_{3}\text{C}_{3}$ exhibit critical temperatures of 102 K and 115 K at 0 and 15 GPa, respectively, potentially setting a new record for the highest $T_\text{c}$ among metal-stuffed B$-$C clathrates. Moreover, several of these high-$T_\text{c}$ compounds are thermodynamically stable under moderate pressure-temperature conditions, comparable to those realized in recent ternary B$-$C clathrate syntheses, indicating their experimental synthesizability.

Our results highlight the necessity of accounting for anharmonic and nuclear quantum effects in accurately predicting the structural and superconducting behavior of lightweight materials. In metal-stuffed B$-$C clathrates, anharmonicity leads to lattice expansion and notable changes in the phonon spectrum, especially in the acoustic branches and key optical modes such as $E_g$ and $E_u$, which exhibit strong electron-phonon coupling and thus significantly influence the predicted superconducting critical temperature. Therefore, capturing the interplay between anharmonic lattice dynamics and electron-phonon interactions is essential for discovering and optimizing lightweight superconductors operable under experimentally accessible conditions.

\section{IV.  METHODS}

\textbf{DF(P)T calculations} 

All DFT and DFPT calculations of electronic and vibrational properties were carried out using the plane-wave pseudopotential code Quantum Espresso\cite{QE}, scalar-relativistic optimized norm-conserving Vanderbilt (ONCV) pseudopotentials\cite{ONCV-1,ONCV-2}, and the PBE-GGA exchange and correlation functional\cite{GGA}. The cutoffs for the wavefunctions and density were chosen as 80 Ry and 640 Ry, respectively. Self-consistent electron density and harmonic phonons were calculated by employing $24 \times 24 \times 24$ $k$-point meshes and $6 \times 6 \times 6$ $q$-point meshes.

\textbf{SSCHA calculations}

Anharmonic effects were estimated using the SSCHA code\cite{sscha-2013,SSCHA-2014,sscha-2017,sscha-2018} by minimizing the free energy with respect to the average atomic positions \textbf{R} and the force constants $\Phi$. We used the DFT equilibrium atomic positions and the DFPT dynamical matrix on a $2 \times 2 \times 2$ $q$-point grid as the initial guesses for \textbf{R} and $\Phi$. 

The total energy, forces, and stress tensors for each configuration were obtained using DFT in Quantum Espresso\cite{QE}, combined with machine learning potentials (MLPs) when necessary, as detailed below. At the end of each minimization run, a new ensemble with a larger number of configurations $N_\text{c}$ is generated from the minimized trial density matrix, iterating until convergence. Two stopping criteria were set for the minimization loop: an effective sample size with a Kong–Liu ratio of 0.6, and a ratio of the free energy gradient to the stochastic error of the auxiliary dynamical matrix below $10^{-6}$. In DFT-based calculations, $N_\text{c}$ was increased to 1000 configurations, while in the MLP framework it was raised to $4 \times 10^{4}$ configurations. 

The anharmonic phonon frequencies were obtained from the positional free energy Hessian including the 4-order terms. The difference between the harmonic and anharmonic dynamical matrices in the $2 \times 2 \times 2$ phonon-momentum grid was interpolated to a $6 \times 6 \times 6$ grid. Adding the harmonic $6 \times 6 \times 6$ grid dynamical matrices to the result, the anharmonic $6 \times 6 \times 6$ $q$-grid dynamical matrices were obtained. 

\textbf{Machine learning potentials}

The machine learning potentials (MLPs) were trained and evaluated using the Attention Coupled Neural Network (ACNN) package. We use Chebyshev polynomial expansions up to the 12th order for radial basis functions within a $6.0$ \AA\ cutoff and up to the 10th order for angular functions within a $5.0$ \AA\ cutoff. 

The MLP was trained on 1000 structures randomly chosen from the last two SSCHA-DFT populations, each containing 1000 structures in a $2 \times 2 \times 2$ supercell. Separate MLPs were trained for each compound to ensure the highest possible accuracy of our calculations. We validated the potentials on the remaining 1000 structures, finding an RMSE of less than 0.5 meV$\cdot$atom$^{-1}$ for the total energy and less than 50 meV$\cdot$\AA$^{-1}$ for the force components. The validations and RMSEs are shown in Supplementary Information.

\textbf{ME theory}

The superconducting gap and $T_\text{c}$ values at both the harmonic and anharmonic levels were calculated by numerically solving the isotropic Eliashberg equations\cite{eliashberg-1960,eliashberg-2007,eliashberg-2018}, using values of $\mu^*$ ranging from 0.1 to 0.15. The Matsubara frequency cutoff is taken to be about 10 times the highest phonon frequency.

\textbf{Anisotropic Eliashberg calculations}

To assess the effect of momentum-dependent electron-phonon coupling (EPC), we performed fully anisotropic Migdal$-$Eliashberg calculations using the EPW code\cite{EPW-1,EPW-2,EPW-3}. The calculations follow the setup described by Gai \textit{et al.}\cite{Rb0.4}. We employed $60 \times 60 \times 60$ electron and $20 \times 20 \times 20$ phonon grids, with Gaussian broadenings of 50~meV and 0.5~meV for the electronic and phononic $\delta$ functions, respectively. The Matsubara summation was truncated at $\omega_c = 1.0$~eV (approximately ten times the maximum phonon frequency), and a 1.0~eV energy window around the Fermi level was used for Wannier interpolation.

The electron-phonon matrix elements were interpolated using maximally localized Wannier functions (MLWFs). For all compounds, 24 initial projections were used, consisting of sp$^3$-hybridized orbitals centered on B–C bonds, and atomic orbitals on the metal atoms (Sr: $p$; K: $s, p$; Rb: $p$; Pb: $s, d$). The resulting MLWFs exhibit good spatial localization, ensuring accurate interpolation of both the band structure and EPC matrix elements. The anisotropic Eliashberg equations were solved self-consistently, following the methodology described in Refs.\cite{EPW-1,EPW-2,EPW-3}.

\section{Code availability}
The SSCHA code (https://github.com/SSCHAcode/python-sscha) is open source and is based on the GNU General Public License v3.0.

The ACNN code is available from Y. X. upon reasonable request. 

\section{Data availability}

All data in the paper are available from the corresponding author upon request.

\section{Acknowledgments}
This work was supported by the National Natural Science Foundation of China (Grant No. 12374008, 12022408, 12304013, 12374009, 12074138, 22131006, 52288102, and 52090024), the Interdisciplinary Integration and Innovation Project of JLU, Fundamental Research Funds for the Central Universities and the Program for JLU Science and Technology Innovative Research Team (JLUSTIRT), open project from state key laboratory of superhard materials (No. 202408), and College Student Innovation and Entrepreneurship Training Program (No. S202310183153). I.E. acknowledges financial support from the European Research Council (ERC) under the European Unions Horizon 2020 research and innovation program (Grant Agreement No. 802533), the Spanish Ministry of Science and Innovation (Grant No. PID2022142861NA-I00), the Department of Education, Universities and Research of the Eusko Jaurlaritza and the University of the Basque Country UPV/EHU (Grant No. IT1527-22), and Simons Foundation through the Collaboration on New Frontiers in Superconductivity (Grant No. SFI-MPS-NFS-00006741- 10). 
Y.-W.F. acknowledges the Extraordinary Grant of CSIC (No. 2025ICT122) and the IKUR Strategy-High Performance Computing and Artificial Intelligence (HPC$\&$AI) 2025-2026 of the Department of Science, Universities and Innovation of the Basque Government.
Technical and human support provided by DIPC Supercomputing Center is gratefully acknowledged. 

\section{Author contributions}
Y.X. designed and supervised the project. W.Z. performed most of the calculations and drafted the manuscript. Y.S. contributed to figure design and assisted with writing and revising the manuscript. J.L. assisted with calculations involving machine-learning potentials. P.Y. performed part of the anharmonic calculations. Y.-W.F. and I.E. provided technical guidance and support for anharmonic calculations. T.I., X.Z., H.Li, and H.Liu contributed to manuscript revision and discussion. All authors have read and approved the final version of the manuscript. W.Z. and Y.S. contributed equally to this work.

\section{Ethics declarations}
\textbf{Competing interests} \\
The authors declare no competing interests.

\section{Additional information}
Materials \& Correspondence should be addressed to Y.X.


\section{References}
\begin{thebibliography}{10}

\bibitem{Overview-2021}
Boeri Lilia, Richard Hennig, Peter Hirschfeld, Gianni Profeta, Antonio Sanna, Eva Zurek, Warren~E Pickett, Maximilian Amsler, Ranga Dias, Mikhail~I Eremets, et~al.
\newblock {The 2021 room-temperature superconductivity roadmap}.
\newblock {\em Journal of Physics: Condensed Matter}, 34(18):183002, 2022.

\bibitem{OverviewofSuperconductivityYing-1}
Ying Sun, Xin Zhong, Hanyu Liu, and Yanming Ma.
\newblock {Clathrate metal superhydrides under high-pressure conditions: enroute to room-temperature superconductivity}.
\newblock {\em National Science Review}, 11(7):nwad270, 2024.

\bibitem{OverviewofSuperconductivityYing-2}
Xiaohua Zhang, Yaping Zhao, Fei Li, and Guochun Yang.
\newblock {Pressure-induced hydride superconductors above 200 K}.
\newblock {\em Matter and Radiation at Extremes}, 6(6), 2021.

\bibitem{Overview-Cui}
Mingyang Du, Wendi Zhao, Tian Cui, and Defang Duan.
\newblock {Compressed superhydrides: the road to room temperature superconductivity}.
\newblock {\em Journal of Physics: Condensed Matter}, 34(17):173001, 2022.

\bibitem{BCS-1}
N.~W. Ashcroft.
\newblock {Metallic Hydrogen: A High-Temperature Superconductor?}
\newblock {\em Physical Review Letters}, 21(26):1748--1749, 1968.

\bibitem{BCS-2}
J.~Bardeen, L.~N. Cooper, and J.~R. Schrieffer.
\newblock {Theory of Superconductivity}.
\newblock {\em Physical Review}, 108(5):1175--1204, 1957.

\bibitem{Belli-2021}
Francesco Belli, Trinidad Novoa, J.~Contreras-Garc{\'i}a, and Ion Errea.
\newblock Strong correlation between electronic bonding network and critical temperature in hydrogen-based superconductors.
\newblock {\em Nature Communications}, 12(1):5381, 2021.

\bibitem{MgB2-original}
Jun Nagamatsu, Norimasa Nakagawa, Takahiro Muranaka, Yuji Zenitani, and Jun Akimitsu.
\newblock {Superconductivity at 39 K in magnesium diboride}.
\newblock {\em Nature}, 410(6824):63--64, 2001.

\bibitem{review-BandCClathrates}
Shoji Yamanaka.
\newblock {Silicon clathrates and carbon analogs: high pressure synthesis, structure, and superconductivity}.
\newblock {\em Dalton Transactions}, 39(8):1901--1915, 2010.

\bibitem{SiandC-clathrates}
Damien Conn{\'e}table and Xavier Blase.
\newblock {Electronic and superconducting properties of silicon and carbon clathrates}.
\newblock {\em Applied Surface Science}, 226(1-3):289--297, 2004.

\bibitem{B-QC}
Anagh Bhaumik, Ritesh Sachan, Siddharth Gupta, and Jagdish Narayan.
\newblock {Discovery of high-temperature superconductivity (T c= 55 K) in B-doped Q-carbon}.
\newblock {\em ACS nano}, 11(12):11915--11922, 2017.

\bibitem{metal-intercalated-graphites2015}
Robert~P Smith, Thomas~E Weller, Christopher~A Howard, Mark~PM Dean, Kaveh~C Rahnejat, Siddharth~S Saxena, and Mark Ellerby.
\newblock {Superconductivity in graphite intercalation compounds}.
\newblock {\em Physica C: Superconductivity and Its Applications}, 514:50--58, 2015.

\bibitem{heavily-boron-doped-diamond}
Jonathan~E. Moussa and Marvin~L. Cohen.
\newblock {Constraints on ${T}_{c}$ for superconductivity in heavily boron-doped diamond}.
\newblock {\em Phys. Rev. B}, 77:064518, Feb 2008.

\bibitem{boron-doped-graphane}
Ya~Cheng, Xianlong Wang, Jie Zhang, Kaishuai Yang, Caoping Niu, and Zhi Zeng.
\newblock {Superconductivity of boron-doped graphane under high pressure}.
\newblock {\em RSC Advances}, 9(14):7680--7686, 2019.

\bibitem{doped-C-clathrates-77k}
F~Zipoli, M~Bernasconi, and G~Benedek.
\newblock {Electron-phonon coupling in halogen-doped carbon clathrates from first principles}.
\newblock {\em Physical Review B-- Condensed Matter and Materials Physics}, 74(20):205408, 2006.

\bibitem{layered-metal-intercalated-borocarbides-Li}
Charlsey~R Tomassetti, Gyanu~P Kafle, Edan~T Marcial, Elena~R Margine, and Aleksey~N Kolmogorov.
\newblock {Prospect of high-temperature superconductivity in layered metal borocarbides}.
\newblock {\em Journal of Materials Chemistry C}, 12(13):4870--4884, 2024.

\bibitem{layered-metal-intercalated-borocarbides-M}
Wataru Hayami, Xavier Rocquefelte, and Jean-Fran{\c{c}}ois Halet.
\newblock {Possible Superconductivity for Layered Metal Boride Carbide Compounds MB2C2 (M= Alkali, Alkaline-Earth, or Rare-Earth Metals)}.
\newblock {\em Inorganic Chemistry}, 2024.

\bibitem{BC5}
Quan Li, Hui Wang, Yongjun Tian, Yang Xia, Tian Cui, Julong He, Yanming Ma, and Guangtian Zou.
\newblock {Superhard and superconducting structures of BC5}.
\newblock {\em Journal of Applied Physics}, 108(2), 2010.

\bibitem{boron-carbon-Material-trends}
Santanu Saha, Simone Di~Cataldo, Maximilian Amsler, Wolfgang Von Der~Linden, and Lilia Boeri.
\newblock {High-temperature conventional superconductivity in the boron-carbon system: Material trends}.
\newblock {\em Physical Review B}, 102(2):024519, 2020.

\bibitem{srbc-prr}
Li~Zhu, Hanyu Liu, Maddury Somayazulu, Yue Meng, Piotr~A. Gu{\'n}ka, Thomas~B. Shiell, Curtis Kenney-Benson, Stella Chariton, Vitali~B. Prakapenka, Hyeok Yoon, Jarryd~A. Horn, Johnpierre Paglione, Roald Hoffmann, R.~E. Cohen, and Timothy~A. Strobel.
\newblock {Superconductivity in ${\mathrm{SrB}}_{3}{\mathrm{C}}_{3}$ clathrate}.
\newblock {\em Physical Review Research}, 5(1):013012, 2023.

\bibitem{srbc-sci}
Li~Zhu, Gustav~M. Borstad, Hanyu Liu, Piotr~A. Gu{\'n}ka, Michael Guerette, Juli-Anna Dolyniuk, Yue Meng, Eran Greenberg, Vitali~B. Prakapenka, Brian~L. Chaloux, Albert Epshteyn, Ronald~E. Cohen, and Timothy~A. Strobel.
\newblock {Carbon-boron clathrates as a new class of sp3-bonded framework materials}.
\newblock {\em Science Advances}, 6(2):eaay8361.

\bibitem{SrBC-BaBC}
Jin-Ning Wang, Xun-Wang Yan, and Miao Gao.
\newblock {High-temperature superconductivity in ${\mathrm{SrB}}_{3}{\mathrm{C}}_{3}$ and ${\mathrm{BaB}}_{3}{\mathrm{C}}_{3}$ predicted from first-principles anisotropic Migdal-Eliashberg theory}.
\newblock {\em Physical Review B}, 103(14):144515, 2021.

\bibitem{Rb0.4}
Ting-Ting Gai, Peng-Jie Guo, Huan-Cheng Yang, Yan Gao, Miao Gao, and Zhong-Yi Lu.
\newblock {Van Hove singularity induced phonon-mediated superconductivity above 77 K in hole-doped ${\mathrm{SrB}}_{3}{\mathrm{C}}_{3}$}.
\newblock {\em Physical Review B}, 105(22):224514, 2022.

\bibitem{SrRb}
Peiyu Zhang, Xue Li, Xin Yang, Hui Wang, Yansun Yao, and Hanyu Liu.
\newblock {Path to high-${T}_{\mathrm{c}}$ superconductivity via Rb substitution of guest metal atoms in the $\mathrm{Sr}{\mathrm{B}}_{3}{\mathrm{C}}_{3}$ clathrate}.
\newblock {\em Physical Review B}, 105(9):094503, 2022.

\bibitem{XBC}
Simone Di~Cataldo, Shadi Qulaghasi, Giovanni~B. Bachelet, and Lilia Boeri.
\newblock {High-${T}_{c}$ superconductivity in doped boron-carbon clathrates}.
\newblock {\em Physical Review B}, 105(6):064516, 2022.

\bibitem{XYBC}
Nisha Geng, Katerina~P. Hilleke, Li~Zhu, Xiaoyu Wang, Timothy~A. Strobel, and Eva Zurek.
\newblock Conventional high-temperature superconductivity in metallic, covalently bonded, binary--guest c--b clathrates.
\newblock {\em Journal of the American Chemical Society}, 145(3):1696--1706, 2023.

\bibitem{BC-NH4}
Ying Sun and Li~Zhu.
\newblock {Hydride Units Filled B--C Clathrate: A New Pathway for High-Temperature Superconductivity at Ambient Pressure}.
\newblock {\em arXiv preprint arXiv:2311.01656}, 2023.

\bibitem{LaBC}
Timothy~A. Strobel, Li~Zhu, Piotr~A. Gu{\'n}ka, Gustav~M. Borstad, and Michael Guerette.
\newblock {A Lanthanum-Filled Carbon--Boron Clathrate}.
\newblock {\em Angewandte Chemie International Edition}, 60(6):2877--2881, 2021.

\bibitem{SSCHA-H3S-1}
Ion Errea, Matteo Calandra, Chris~J. Pickard, Joseph~R. Nelson, Richard~J. Needs, Yinwei Li, Hanyu Liu, Yunwei Zhang, Yanming Ma, and Francesco Mauri.
\newblock {Quantum hydrogen-bond symmetrization in the superconducting hydrogen sulfide system}.
\newblock {\em Nature}, 532(7597):81--84, 2016.

\bibitem{SSCHA-LaH10}
Ion Errea, Francesco Belli, Lorenzo Monacelli, Antonio Sanna, Takashi Koretsune, Terumasa Tadano, Raffaello Bianco, Matteo Calandra, Ryotaro Arita, Francesco Mauri, and Jos{\'e}~A. Flores-Livas.
\newblock {Quantum crystal structure in the 250-kelvin superconducting lanthanum hydride}.
\newblock {\em Nature}, 578(7793):66--69, 2020.

\bibitem{SSCHA-program}
Lorenzo Monacelli, Raffaello Bianco, Marco Cherubini, Matteo Calandra, Ion Errea, and Francesco Mauri.
\newblock {The stochastic self-consistent harmonic approximation: calculating vibrational properties of materials with full quantum and anharmonic effects}.
\newblock {\em Journal of Physics: Condensed Matter}, 33(36):363001, 2021.

\bibitem{sscha-2013}
Ion Errea, Matteo Calandra, and Francesco Mauri.
\newblock {First-Principles Theory of Anharmonicity and the Inverse Isotope Effect in Superconducting Palladium-Hydride Compounds}.
\newblock {\em Phys. Rev. Lett.}, 111:177002, Oct 2013.

\bibitem{SSCHA-2014}
Ion Errea, Matteo Calandra, and Francesco Mauri.
\newblock {Anharmonic free energies and phonon dispersions from the stochastic self-consistent harmonic approximation: Application to platinum and palladium hydrides}.
\newblock {\em Phys. Rev. B}, 89:064302, Feb 2014.

\bibitem{sscha-2017}
Raffaello Bianco, Ion Errea, Lorenzo Paulatto, Matteo Calandra, and Francesco Mauri.
\newblock {Second-order structural phase transitions, free energy curvature, and temperature-dependent anharmonic phonons in the self-consistent harmonic approximation: Theory and stochastic implementation}.
\newblock {\em Phys. Rev. B}, 96:014111, Jul 2017.

\bibitem{sscha-2018}
Lorenzo Monacelli, Ion Errea, Matteo Calandra, and Francesco Mauri.
\newblock {Pressure and stress tensor of complex anharmonic crystals within the stochastic self-consistent harmonic approximation}.
\newblock {\em Phys. Rev. B}, 98:024106, Jul 2018.

\bibitem{eliashberg-1960}
GM~Eliashberg.
\newblock {Interactions between electrons and lattice vibrations in a superconductor}.
\newblock {\em Sov. Phys. JETP}, 11(3):696--702, 1960.

\bibitem{eliashberg-2007}
Feliciano Giustino, Marvin~L Cohen, and Steven~G Louie.
\newblock {Electron-phonon interaction using Wannier functions}.
\newblock {\em Physical Review B-- Condensed Matter and Materials Physics}, 76(16):165108, 2007.

\bibitem{eliashberg-2018}
Antonio Sanna, Jos{\'e}~A Flores-Livas, Arkadiy Davydov, Gianni Profeta, Kay Dewhurst, Sangeeta Sharma, and EKU Gross.
\newblock {Ab initio Eliashberg theory: making genuine predictions of superconducting features}.
\newblock {\em Journal of the Physical Society of Japan}, 87(4):041012, 2018.

\bibitem{Bader}
Graeme Henkelman, Andri Arnaldsson, and Hannes J{\'o}nsson.
\newblock A fast and robust algorithm for bader decomposition of charge density.
\newblock {\em Computational Materials Science}, 36(3):354--360, 2006.

\bibitem{COHP-0}
Richard Dronskowski and Peter~E. Bloechl.
\newblock Crystal orbital hamilton populations (cohp): energy-resolved visualization of chemical bonding in solids based on density-functional calculations.
\newblock {\em The Journal of Physical Chemistry}, 97(33):8617--8624, 1993.

\bibitem{COHP-1}
Volker~L. Deringer, Andrei~L. Tchougr{\'e}eff, and Richard Dronskowski.
\newblock Crystal orbital hamilton population (cohp) analysis as projected from plane-wave basis sets.
\newblock {\em The Journal of Physical Chemistry A}, 115(21):5461--5466, 2011.

\bibitem{Mulliken}
Christina Ertural, Simon Steinberg, and Richard Dronskowski.
\newblock Development of a robust tool to extract mulliken and l{\"o}wdin charges from plane waves and its application to solid-state materials.
\newblock {\em RSC Adv.}, 9:29821--29830, 2019.

\bibitem{lobster}
Stefan Maintz, Volker~L. Deringer, Andrei~L. Tchougr{\'e}eff, and Richard Dronskowski.
\newblock Lobster: A tool to extract chemical bonding from plane-wave based dft.
\newblock {\em Journal of Computational Chemistry}, 37(11):1030--1035, 2016.

\bibitem{Annealed-importance-sampling2001}
Radford~M Neal.
\newblock {Annealed importance sampling}.
\newblock {\em Statistics and Computing}, 11:125--139, 2001.

\bibitem{Monte-Carlo-1994}
R~Marto{\v{n}}{\'a}k and E~Tosatti.
\newblock {Path-integral Monte Carlo study of a model two-dimensional quantum paraelectric}.
\newblock {\em Physical Review B}, 49(18):12596, 1994.

\bibitem{Tiago-AFM2024}
Tiago F.~T. Cerqueira, Yue-Wen Fang, Ion Errea, Antonio Sanna, and Miguel A.~L. Marques.
\newblock {Searching Materials Space for Hydride Superconductors at Ambient Pressure}.
\newblock {\em Advanced Functional Materials}, 34(40):2404043, 2024.

\bibitem{DAC}
Simone Anzellini and Silvia Boccato.
\newblock A practical review of the laser-heated diamond anvil cell for university laboratories and synchrotron applications.
\newblock {\em Crystals}, 10(6), 2020.

\bibitem{QE}
Paolo Giannozzi, Stefano Baroni, Nicola Bonini, Matteo Calandra, Roberto Car, Carlo Cavazzoni, Davide Ceresoli, Guido~L Chiarotti, Matteo Cococcioni, Ismaila Dabo, et~al.
\newblock {QUANTUM ESPRESSO: a modular and open-source software project for quantum simulations of materials}.
\newblock {\em Journal of Physics: Condensed Matter}, 21(39):395502, 2009.

\bibitem{ONCV-1}
Martin Schlipf and Fran{\c{c}}ois Gygi.
\newblock {Optimization algorithm for the generation of ONCV pseudopotentials}.
\newblock {\em Computer Physics Communications}, 196:36--44, 2015.

\bibitem{ONCV-2}
D.~R. Hamann.
\newblock {Optimized norm-conserving Vanderbilt pseudopotentials}.
\newblock {\em Phys. Rev. B}, 88:085117, Aug 2013.

\bibitem{GGA}
John~P Perdew, Kieron Burke, and Matthias Ernzerhof.
\newblock {Generalized gradient approximation made simple}.
\newblock {\em Physical review letters}, 77(18):3865, 1996.

\bibitem{EPW-1}
Hyoung~Joon Choi, Marvin~L. Cohen, and Steven~G. Louie.
\newblock Anisotropic eliashberg theory of mgb2: Tc, isotope effects, superconducting energy gaps, quasiparticles, and specific heat.
\newblock {\em Physica C: Superconductivity}, 385(1):66--74, 2003.

\bibitem{EPW-2}
E.~R. Margine and F.~Giustino.
\newblock Anisotropic migdal-eliashberg theory using wannier functions.
\newblock {\em Phys. Rev. B}, 87:024505, Jan 2013.

\bibitem{EPW-3}
S.~Ponc{\'e}, E.~R. Margine, C.~Verdi, and F.~Giustino.
\newblock Epw: Electron\textendash phonon coupling, transport and superconducting properties using maximally localized wannier functions.
\newblock {\em Computer Physics Communications}, 209:116--133, 2016.

\end{thebibliography}
\end{document}